\documentclass[aps,prl,reprint]{revtex4-2}

\usepackage{jltdefs}
\graphicspath{{figs/}}

\usepackage[bbgreekl]{mathbbol} % for \bbsigma

\usepackage{chngcntr}
\counterwithout{equation}{section}

\begin{document}

%
% Notation
%

% For reasons I don't understand, if I use a two-column mode I have to
% generate some output before the nomenclature commands will "take".
\phantom{}

% General definitions (not in notation table)
\renewcommand{\d}{\mathrm{d}}
\mathnotation{\rot}{{\mathrm{r}}}
\mathnotation{\eff}{{\mathrm{eff}}}

% Show in order of introduction in the paper, for nomenclature.

\mathnotation[particle position]{\xv}{\bm{\xc}}
\mathnotationentry[swimming angle]{\phi}
\renewmathnotation[time]{\t}{t}

% Unit vector "hat" notation conflicts with direct sum.
% These are defined in jltdefs, so overwrite here.
\renewcommand{\ruv}{\bm{p}_\parallel}
\renewcommand{\phiuv}{\bm{p}_\perp}
\mathnotationentry[swimming direction]{\ruv}
\mathnotationentry[perpendicular to swimming direction]{\phiuv}

\mathnotation{\Uc}{U}
\mathnotation[swimming velocity $\Uc\ruv$]{\Uv}{\bm{U}}
\mathnotationentry[angular swimming speed]{\Omega}

% Spatial diffusion.
\mathnotation{\Dc}{D}
\mathnotation{\Dbb}{\mathbb{\Dc}}
\mathnotation[$\parallel$ translational diffusivity]{\DX}{\Dc_\parallel}
\mathnotation[$\perp$ translational diffusivity]{\DY}{\Dc_\perp}
\mathnotation[rotational diffusivity]{\Dr}{\Dc_\rot}

% Wiener process
\mathnotation{\wc}{w}
\mathnotationentry[standard Wiener process, %
  $i \in \{\parallel,\perp,\rot\}$]{\wc_i}%
\mathnotation{\dotwc}{\dot{\wc}}
\mathnotation{\dotwv}{\dot{\bm{\wc}}}
\mathnotation{\dotwcX}{\dotwc_\parallel}
\mathnotation{\dotwcY}{\dotwc_\perp}
\mathnotation{\dotwcr}{\dotwc_\rot}

\mathnotationentry[Inverse temperature $1/k_{\mathrm{B}}T$]{\beta}

% Force
\mathnotation{\fc}{f}
\mathnotation[propulsion force acting on particle]{\fv}{\bm{\fc}}
\mathnotation[force noise in the $\parallel$ direction]{\EX}{\Ec_\parallel}
\mathnotation[force noise in the $\perp$ direction]{\EY}{\Ec_\perp}
\mathnotation{\Fc}{F}
\mathnotation{\FX}{\Fc_\parallel}
\mathnotation{\FY}{\Fc_\perp}
\mathnotation[mean force $\FX\,\ruv + \FY\,\phiuv$]{\Fv}{\bm{\Fc}}
\mathnotationentry[torque $\ell(\FY + \sqrtEY\,\phiuv)$]{\tau}
\mathnotationentry[$\fv$ acts at $\ell\ruv$ from center of reaction]{\ell}

\mathnotation[mass of particle]{\mass}{m}
\mathnotation[moment of inertia about $\ruv\times\phiuv$]{\Inerc}{I}
\mathnotation{\uc}{u}
\mathnotation[particle velocity]{\uv}{\bm{\uc}}
\mathnotationentry[angular speed]{\omega}
\mathnotation[$\parallel$ component of resistance]{\sigmaX}{\sigma_\parallel}
\mathnotation[$\perp$ component of resistance]{\sigmaY}{\sigma_\perp}
\mathnotation[rotational resistance]{\sigmar}{\sigma_\rot}
\mathnotation{\Kc}{K}
\mathnotation[resistance matrix %
  $\Qbb\cdot\diag(\sigmaX,\sigmaY)\cdot\Qbb^\T$]{\Kbb}{\mathbb{\Kc}}
\mathnotation{\Qc}{Q}
\mathnotation[2D CCW rotation by~$\phi$]{\Qbb}{\mathbb{\Qc}}
\mathnotationentry[block-diagonal matrix]{\diag}

% hat quantities
\mathnotation[$(\xv,\phi)$]{\hxv}{\widehat{\xv}}
\mathnotation[$(\uv,\omega)$]{\huv}{\widehat{\uv}}
\mathnotation[$(\Uv,\Omega)$]{\hUv}{\widehat{\Uv}}
\mathnotation{\Sigmabb}{\mathbb{\Sigma}}
\mathnotation[noise coupling matrix]{\hSigmabb}{\widehat{\Sigmabb}}
\mathnotationentry[$(\wc_\parallel,\wc_\perp)$]{\bm{\wc}}
\mathnotation{\Bc}{B}
\renewmathnotation{\Bbb}{\mathbb{\Bc}}
\mathnotation[$\diag(\Kbb/\mass,\sigmar/\Inerc)$]%
  {\hBbb}{\widehat{\Bbb}}

\mathnotation[probability density $\pp(\hxv,\huv,\t)$]{\pp}{p}
\mathnotationentry[expansion parameter (overdamped)]{\eps}
\renewmathnotation[collision operator]{\L}{\mathcal{L}}
\mathnotation{\Ec}{E}
\mathnotation{\Ebb}{\mathbb{{\Ec}}}
\mathnotation[grand velocity noise matrix]{\hEbb}{\widehat{\Ebb}}
\mathnotationentry[outer product $(\bm{a}\otimes\bm{b})_{ij}=a_ib_j$]{\otimes}
\mathnotationentry[Gaussian distribution $\varphi(\hxv,\huv)$]{\varphi}
\mathnotation{\Ac}{A}
\mathnotation{\Abb}{\mathbb{\Ac}}
\mathnotation[grand covariance matrix $\hAbb(\hxv)$]{\hAbb}{\widehat{\Abb}}

\mathnotation[matrix transpose superscript]{\T}{\top}
\mathnotation[rotation about third axis]{\hQbb}{\widehat{\Qbb}}

\mathnotation[prob.\ density $\PP(\hxv,\t)$ (overdamped)]{\PP}{P}
\mathnotation{\chiv}{\bm{\chi}}
\mathnotation[`cell problem' functions]{\hchiv}{\widehat{\bm{\chi}}}

\mathnotationentry[adjoint of $\L$]{\L^*}

\mathnotation{\Vc}{V}
\mathnotation[noise-induced drift]{\Vv}{\bm{\Vc}}
\mathnotation{\hDc}{\widehat{\Dc}}
\mathnotation[grand diffusion matrix]{\hDbb}{\widehat{\Dbb}}

\mathnotation[particle size]{\aaa}{a}
\mathnotation[P\'eclet number $\nofrac{\lvert\Vv\rvert\aaa}{\DY}$]%
  {\PeY}{\Pe_\perp}
\mathnotation[P\'eclet number $\nofrac{\lvert\Vv\rvert}{\Dr\aaa}$]%
  {\Per}{\Pe_\rot}

\mathnotation{\Wc}{W}
\mathnotation[total velocity $\Uv + \Vv$]{\Wv}{\bm{\Wc}}

\mathnotationentry[expansion parameter (large-scale)]{\delta}
\mathnotation{\Xc}{X}
\mathnotation[large-scale variable $\sim \delta^{-1}$]{\Xv}{\bm{\Xc}}
\mathnotation[long-time variable $\sim \delta^{-2}$]{\Time}{T}
\mathnotation[prob.\ density $\PPt(\Xv,\Time)$ (large scale)]%
  {\PPt}{\mathcal{P}}

\mathnotation[effective diffusivity $\tfrac12(\DX + \DY) + \Dt$]%
  {\Deff}{\Dc_\eff}
\mathnotation[added diffusivity]{\Dt}{\widetilde{\Dc}}
\mathnotationentry[added diffusivity for non-swimmer]{\Dt_0}

\mathnotation[double dot product $\mathbb{A}\cddot\mathbb{B} = A_{ij}B_{ij}$]%
  {\cddot}{:} %
\mathnotation[triple dot product $\mathbb{A}\cdddot\mathbb{B} = %
  A_{ijk}B_{ijk}$]{\cdddot}{\,\raisebox{-.2em}{$\vdots$}\,}

% Convenient defs
\mathnotation{\sqrtDX}{\sqrt{\smash[b]{2\DX}}}
\mathnotation{\sqrtDY}{\sqrt{\smash[b]{2\DY}}}
\mathnotation{\sqrtDr}{\sqrt{\smash[b]{2\Dr}}}
\mathnotation{\sqrtEX}{\sqrt{\smash[b]{2\EX}}}
\mathnotation{\sqrtEY}{\sqrt{\smash[b]{2\EY}}}

\title{%
  Anisotropic active Brownian particle
  with a fluctuating propulsion force}

\author{Jean-Luc Thiffeault}
\affiliation{Department of Mathematics, University of Wisconsin -- Madison,
  Madison, WI 53706, USA.}
\email{jeanluc@math.wisc.edu}

\author{Jiajia Guo}
\affiliation{Department of Mathematics, University of Michigan, Ann Arbor, MI
  48109, USA}
%

%\date{\today}

\begin{abstract}
  The active Brownian particle (ABP) model describes a swimmer, synthetic or
  living, whose direction of swimming is a Brownian motion.  The swimming is
  due to a propulsion force, and the fluctuations are typically thermal in
  origin.  We present a 2D model where the fluctuations arise from nonthermal
  noise in a propelling force acting at a single point, such as that due to a
  flagellum.  We take the overdamped limit and find several modifications to
  the traditional ABP model.  Since the fluctuating force causes a fluctuating
  torque, the diffusion tensor describing the process has a coupling between
  translational and rotational degrees of freedom.  An anisotropic particle
  also exhibits a mass-dependent noise-induced drift, which does not disappear
  in the overdamped limit.  We show that these effects have measurable
  consequences for the long-time diffusivity of active particles, in
  particular adding a contribution that is independent of where the force
  acts.
\end{abstract}

\maketitle

\noindent
\textit{[Note: at the end of the paper is a Nomenclature of mathematical
  symbols, in the order introduced.]}

\smallskip

Modeling swimming microorganisms is a challenge, since biological entities
resist a simple, uniform description.  Nevertheless we need models to explain
physical observations and develop intuition, and the hope is that the models
capture some essential aspect of an organism's behavior.  For microswimmers,
most modeling efforts impose some randomness to the motion.  The simplest
approach is to use a fixed propulsion speed, together with a random
re-orientation mechanism.  The random re-orientation comes in two main
flavors: a run-and-tumble process where the organism makes large excursions
and changes its orientation sporadically~\cite{Subramanian2009, Nash2010,
  Martens2012, Cates2013, Elgeti2015, Ezhilan2015b, Lee2019}, and a Brownian
process where the direction of swimming gradually varies~\cite{Peruani2007,
  vanTeeffelen2008, Baskaran2008, Romanczuk2011, Romanczuk2012,
  Kurzthaler2016, Kurzthaler2017}.  Both of these models have their place, but
in this letter we focus primarily on the latter, Brownian approach.

\begin{figure}
  \begin{center}
    \includegraphics[width=.6\columnwidth]{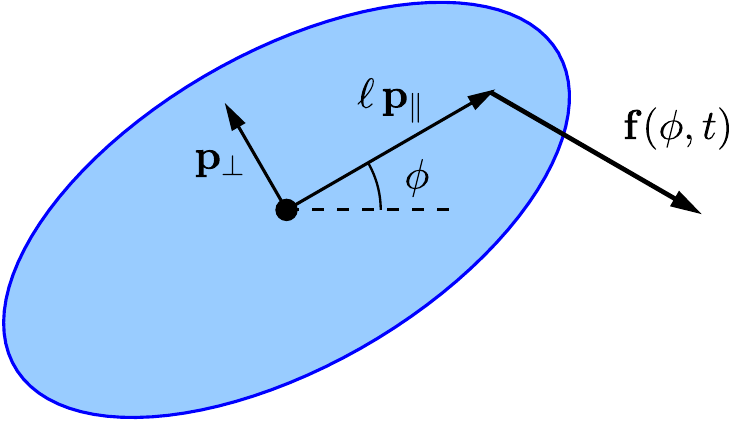}
  \end{center}
  \caption{A 2D particle with orientation~$\phi$ subject to a time-dependent
    force~$\fv$ acting at a point~$\ell\ruv(\phi)$ with respect to its center
    of reaction.}
  \label{fig:activepart}
\end{figure}

In this letter we present a simple model of random microorganism motion where
the swimmer is propelled by a fluctuating force acting at a point
(\cref{fig:activepart}).  The randomness is built into the force as a
covariance matrix, and is not due to interactions with the medium (though such
interactions could be included as well).  Our goal is to derive effective
equations of motion for this simple configuration, which is meant to
represented an organism with a single flagellum.  The resulting equations have
some points of commonality with the well-known Active Brownian particle model
(ABP), but differ in crucial ways.  In particular, there is an inherent
coupling between translational and rotation diffusivities.  In addition, there
is a noise-induced drift that is present regardless of which stochastic
interpretation (It\^o or Stratonovich) is used.

The stochastic equations (SDEs) for the 2D ABP model are \cite{Peruani2007,
  vanTeeffelen2008, Baskaran2008, Romanczuk2011, Romanczuk2012,
  Kurzthaler2016, Kurzthaler2017}
\begin{subequations}
  \label{eq:ABP_standard}
  \begin{align}
    \dot{\xv} &=
    (\Uc + \sqrtDX\,\dotwcX)\,\ruv
    + \sqrtDY\,\phiuv\,\dotwcY\,,
    \\
    \dot{\phi} &= \Omega + \sqrtDr\,\dotwcr\,.
  \end{align}
\end{subequations}
The swimmer is moving at constant speed~$\Uc$ in the direction~$\ruv(\phi)$
and rotating at constant angular speed~$\Omega$.  The translational
noises~$\sqrtDX\,\dotwcX$ and~$\sqrtDY\,\dotwcY$ are respectively along
($\ruv$) and perpendicular ($\phiuv$) to the direction of swimming, and the
rotational noise~$\sqrtDr\,\dotwcr$ affects the swimming direction.
The~$\wc_i(\t)$ are independent standard Wiener processes.
\Cref{eq:ABP_standard} has been very successful in modeling the swimming and
collective behavior of many microorganisms \cite{Ai2013, Solon2015, Zottl2016,
  Wagner2017, Redner2013, Stenhammar2014, ChenThiffeault2021}.  The noises are
often taken to be due to thermal fluctuations, in which case they satisfy the
Einstein--Smoluchowski relations $\Dc_i = (\beta\sigma_i)^{-1}$, for
$i \in \{\parallel,\perp,\rot\}$, where~$\sigma_i$ are the components of the
diagonal grand resistance tensor and~$\beta$ is the inverse
temperature~$1/k_{\mathrm{B}}T$.

In this letter we derive a modified ABP model by assuming that the noise is
due to a fluctuating propulsion force acting at a single point on the
particle, rather than a thermal bath.  We will find several new effects: a new
noise-induced drift term, as well as a diffusion matrix that couples the
rotational and angular degrees of freedom.

A particle subjected to a fluctuating
force~$\fv(\phi,\t) = (\FX + \sqrtEX\,\dotwcX)\,\ruv + (\FY +
\sqrtEY\,\dotwcY)\,\phiuv$ acting at the point~$\ell\,\ruv$ with respect to
the center of reaction~\cite{HappelBrenner} obeys the Langevin equations
\begin{align}
  \label{eq:SDE_inertia}
    \mass\dot{\uv} =
    -\Kbb\cdot\uv + \fv,
    \quad
    \Inerc\,\dot{\omega} = -\sigmar\,\omega + \tau,
\end{align}
where~$\mass$ is the mass, $\Inerc$ the moment of inertia, $\uv$ the velocity,
$\omega$ the angular velocity,
and~$\Kbb = \Qbb\cdot\diag(\sigmaX,\sigmaY)\cdot\Qbb^\T$ the resistance
matrix, with~$\Qbb(\phi)$ a~$2\times2$ rotation matrix.  The force exerts a
torque~$\tau(\t)=\ell\,(\FY + \sqrtEY\,\dotwcY)$~\footnote{We assume for
  simplicity that the center of mass coincides with the center of reaction.}.

A brief note on the validity of~\cref{eq:SDE_inertia} is in order.  We follow
many authors such as~\cite{Kramer2004,Delong2015} and use a linear damping law
in \cref{eq:SDE_inertia}, which as first pointed out by
Lorentz~\cite{Lorentz_volV} is strictly only valid in the limit where the
fluid density is less than the particle density~\cite{Hauge1973, Hinch1975,
  Durr1981, Roux1992, Bocquet1997, Donev2014}.  The theory could be extended
to allow for a memory kernel, the so-called Basset--Boussinesq integral
term~\cite{Basset1888,Boussinesq1903}, but then the process is non-Markovian
and we cannot recover a simple Fokker--Planck equation as detailed below.
Nevertheless, we expect that this memory effect is unlikely to \emph{decrease}
correlations, and so the effects presented here might be modified but would
not disappear.

We rewrite the system~\eqref{eq:SDE_inertia} in the standard form
\begin{align}
  \frac{\d\hxv}{\d\t}
  =
  \huv,
  \quad
  \frac{\d\huv}{\d\t}
  =
  \hBbb\cdot(\hUv - \huv) + \hSigmabb\cdot\dotwv
  \label{eq:standard}
\end{align}
where~$\hxv = (\xv,\phi)$, $\huv = (\uv,\omega)$,
$\dotwv = (\dotwcX,\dotwcY)$, $\hBbb = \diag(\Kbb/\mass,\sigmar/\Inerc)$,
$\hUv = (\Uv,\Omega) = (\Kbb^{-1}\cdot\Fv,\ell\FY/\sigmar)$, and
\begin{equation}
  \hSigmabb
  =
  \begin{pmatrix}
    (\sqrtEX/\mass)\,\ruv & (\sqrtEY/\mass)\,\phiuv \\
    0 & \sqrtEY\,\ell/\Inerc
  \end{pmatrix}.
\end{equation}
The third components of hat-wearing vectors and matrices pertain to angular
quantities.

Typically, in the overdamped limit (small mass, or large drag) the
term~$\nofrac{\d\huv}{\d\t}$ in~\eqref{eq:standard} is neglected, resulting in
the equation
\begin{equation}
  \frac{\d\hxv}{\d\t} = \hUv + \hBbb^{-1}\cdot\hSigmabb\cdot\dotwv.
  \label{eq:overdamped_simplistic}
\end{equation}
This recovers something close to the standard ABP
model~\eqref{eq:ABP_standard}, except that here there are only two rather than
three independent noises: the rotational noise is correlated to the
translational noise, since the former is caused by the torque of the latter.
We will see the consequences of this correlation below.

% We can see the noise is additive by computing the Stratonovich drift using
% the full 6x2 noise coupling tensor.
But first note that taking the overdamped limit in this way is suspicious.
The underdamped equations~\eqref{eq:standard} have the same form independent
of the interpretation given to the stochastic product (\ie, It\^o or
Stratonovich), even though the noise appears multiplicative at first glance.
However, the noise coupling matrix~$\hBbb^{-1}\cdot\hSigmabb$
in~\cref{eq:overdamped_simplistic} leads to a nonvanishing drift term when the
stochastic product is interpreted in the Stratonovich
sense~\cite[p.~83]{Oksendal}.  This suggests
that~\cref{eq:overdamped_simplistic} has a uniquely-defined noise-induced
drift term~\cite{Kupferman2004, Lau2007, Farago2017}, but the naive way of
passing from~\eqref{eq:standard} to~\eqref{eq:overdamped_simplistic} does not
tell us what form it should take.

A more systematic approach is required to find the missing noise-induced drift
term~\cref{eq:overdamped_simplistic}.  Instead working with SDEs, we take the
overdamped limit of the Fokker--Planck equation for the probability
density~$\pp(\hxv,\huv,\t)$ corresponding to \cref{eq:standard} (see
\cite{Kupferman2004, Bo2013, Pavliotis, Hottovy2015} for an SDE approach):
\begin{align}
  \eps^2\,\pd_t\pp
  + \eps\,\grad_{\hxv}\cdot(\huv\,\pp)
  + \eps\,\grad_{\huv}\cdot(\hBbb\cdot\hUv\pp)
  = \L\pp
  \label{eq:FP}
\end{align}
where~$\eps$ is a formal expansion parameter, with~$\eps\rightarrow0$ the
overdamped limit, and
\begin{align}
  \L\pp \ldef \grad_{\huv}\cdot\bigl(\hBbb\cdot\huv\,\pp\bigr)
  + \grad_{\huv}\otimes\grad_{\huv}\cddot\bigl(\hEbb\,\pp\bigr)
\end{align}
with \hbox{$\hEbb \ldef \tfrac12\hSigmabb\cdot\hSigmabb^\T$}~\footnote{The
  tensor~$\hEbb$ has zero determinant, indicative of a degenerate parabolic
  problem since there are fewer noises than equations.  In practice this is
  inconsequential, since we can add a bit of thermal noise to remove the
  degeneracy.}%
%We have assumed the It\^{o} interpretation of the SDE, but all
%interpretations are the same since~$\hEbb$ is not a function of~$\huv$.
.  The parameter~$\eps$ expresses the long-time and large-scale rescalings
of~$t$ and~$\hxv$ for which the~$\huv$ degrees of freedom equilibrate.

Now we proceed order-by-order with an
expansion~$\pp = \pp_0 + \eps\,\pp_1 + \cdots$.  At leading order we
have~$\L\pp_0=0$, with solution~$\pp_0 = \PP(\hxv,\t)\,\varphi(\hxv,\huv)$,
where~$\PP$ is yet to be determined and~$\varphi(\hxv,\huv)$ is the invariant
density for an Ornstein--Uhlenbeck process \cite{Riksen}:
\begin{equation}
  \varphi
  =
  (2\pi)^{-3}(\det\hAbb)^{-1/2}\,
  \exp\bigl(-\tfrac12\,\huv\cdot\hAbb^{-1}\cdot\huv
  \bigr).
  \label{eq:ppMB}
\end{equation}
Here the symmetric positive-definite matrix~$\hAbb(\hxv)$ is the unique
solution to the continuous-time Lyapunov equation~\footnote{The solution of
  this matrix problem is implemented as \textbf{LyapunovSolve} in Mathematica,
  \textbf{sylvester} in Matlab, and
  \textbf{scipy.linalg.solve\_continuous\_lyapunov} in Python.}
\begin{equation}
  \hBbb\cdot\hAbb + \hAbb\cdot\hBbb^\T = 2\hEbb
  \label{eq:hBAAB}
\end{equation}
where in our case~$\hBbb = \hBbb^\T$.  When~$\hBbb$ commutes with~$\hEbb$, as
occurs for thermal fluctuations, the solution to~\eqref{eq:hBAAB}
is~$\hAbb = \hEbb\cdot\hBbb^{-1}$; this is not the case here, and we find
instead
\begin{equation}
  \hAbb = \hQbb\cdot
  \begin{pmatrix}
    \frac{\EX}{\mass\sigmaX} & 0 & 0 \\
    0 & \frac{\EY}{\mass\sigmaY} &
    \frac{2\EY\ell}{\mass\sigmar + \Inerc\sigmaY} \\
    0 & \frac{2\EY\ell}{\mass\sigmar + \Inerc\sigmaY} &
    \frac{\EY\ell^2}{\Inerc\sigmar}
  \end{pmatrix}
  \cdot\hQbb^\T
\end{equation}
where~$\hQbb(\phi) = \diag(\Qbb,1)$ is a~$3\times3$ rotation matrix about the
third axis.

At the next order in~$\eps$, we
have~$\L\pp_1 = \grad_{\hxv}\cdot(\huv\,\varphi\,\PP) -
\huv\cdot\hAbb^{-1}\cdot\hBbb\cdot\hUv\varphi\PP$.  The solution can be
written in two pieces~$\pp_1 = \pp_1^{(1)} + \pp_1^{(2)}$, with
$\pp_1^{(1)} = (\grad_{\hxv}\PP - \hUv\cdot\hBbb^\T\cdot\hAbb^{-1}\,\PP)
\cdot\hchiv^{(1)}$ and
$\pp_1^{(2)} = - \tfrac12 \PP\,\grad_{\hxv}\hAbb^{-1} \cdddot\hchiv^{(2)}$,
where~$\hchiv^{(1)}$ and~$\hchiv^{(2)}$ satisfy
\begin{equation}
  \L\hchiv^{(1)} = \huv\,\varphi,
  \qquad
  \L\hchiv^{(2)} = \huv\huv\huv\,\varphi.
  \label{eq:Lchi}
\end{equation}
It is easy to solve
for~$\hchiv^{(1)} = -\hAbb\cdot\hBbb^{-\T}\cdot\hAbb^{-1}\cdot\huv\,\varphi$;
$\hchiv^{(2)}$ is harder to solve for in general.  However, we shall not need
its precise expression in our derivation.

At the next and final order in~$\eps$ we get from \cref{eq:FP}
$\L\pp_2 = \grad_{\hxv}\cdot(\huv\,\pp_1) +
\grad_{\huv}\cdot(\hBbb\cdot\hUv\pp_1) + \pd_\t \pp_0$, to which we need only
apply a solvability condition by integrating over~$\huv$ (denoted by angle
brackets):
\begin{equation}
  \pd_\t\PP
  =
  -
  \grad_{\hxv}\cdot\langle\huv\pp_1\rangle.
  \label{eq:eps1_solv}
\end{equation}
To evaluate the average~$\langle\huv\,\pp_1\rangle$, first note that the
adjoint to~$\L$ is
\begin{equation}
  \L^*g = -\huv\cdot\hBbb^\T\cdot\grad_{\huv}g
  +
  \hEbb:\grad_{\huv}\otimes\grad_{\huv} g
\end{equation}
which satisfies~$\langle g \L f\rangle = \langle (\L^*g) f\rangle$ for
functions~$f$ and~$g$ vanishing as~$\lvert\huv\rvert \rightarrow \infty$.
Multiplying the~$\hchiv^{(1)}$ equation in~\eqref{eq:Lchi} by~$\huv$, we have
\begin{align}
  \langle \huv\,\L\hchiv^{(1)} \rangle
  =
  \langle \huv\huv\,\varphi \rangle
  =
  \hAbb\,.
  \label{eq:YLchi1}
\end{align}
But then using the adjoint property in~\eqref{eq:YLchi1} gives
\begin{align*}
  \langle (\L^*\huv)\,\hchiv^{(1)} \rangle
  =
  \langle (-\hBbb\cdot\huv)\,\hchiv^{(1)} \rangle
  =
  -\hBbb\cdot\langle \huv\,\hchiv^{(1)} \rangle
\end{align*}
from which we
obtain~$\langle \huv \hchiv^{(1)} \rangle = -\hBbb^{-1}\cdot\hAbb$.  We can
play a similar trick with the~$\hchiv^{(2)}$ equation to obtain
$\langle \huv\,\hchiv^{(2)} \rangle = -\hBbb^{-1}\cdot\langle
\huv\huv\huv\huv\,\varphi \rangle$, where the fourth moment for the
Gaussian~$\varphi$ is easily obtained.  We have thus evaluated the required
average~$\langle \huv\,\hchiv^{(2)} \rangle$ without needing to solve
for~$\hchiv^{(2)}$.

After a lengthy but straightforward calculation we find
$\langle\huv\pp_1\rangle = \hUv\PP - \grad_{\hxv}\cdot(\hAbb\,\PP)
\cdot\hBbb^{-\T}$, which we insert back into \eqref{eq:eps1_solv}to finally
obtain
\begin{equation}
  \pd_\t\PP
  +
  \grad_{\hxv}\cdot(\hUv\PP)
  =
  \grad_{\hxv}\cdot(\grad_{\hxv}\cdot(\hAbb\,\PP)
  \cdot\hBbb^{-\T}).
  \label{eq:the_final_PDE0}
\end{equation}
We rewrite~\eqref{eq:the_final_PDE0} in a more convenient form and obtain the
first main result of this letter:
\begin{multline}
  \pd_\t\PP
  +
  \grad_{\xv}\cdot((\Uv+\Vv)\PP)
  +
  \pd_\phi(\Omega\,\PP) \\
  =
  \grad_{\hxv}\otimes\grad_{\hxv}\cddot(\hDbb\,\PP)
  \label{eq:the_final_final_PDE}
\end{multline}
where the noise-induced drift~\cite{Grassia1995, Lau2007, Hottovy2012,
  Hottovy2012b, Hottovy2015, Volpe2016, Farago2017} is
\begin{align}
  \Vv &=
  \frac{2\ell\EY(\sigmaX^{-1} - \sigmaY^{-1})}
  {\sigmar(1 + \Inerc\sigmaY/\mass\sigmar)}
  \,\ruv
  \label{eq:Vv}
\end{align}
% Stratonovich drift of {eq:overdamped_simplistic}, for comparison:
%\begin{align}
%  \Vv_{\text{Strat}} &=
%  -\frac{\ell\EY}{\sigmar\sigmaY}\,\ruv
%\end{align}
and the translational-rotational grand diffusion tensor is
\begin{equation}
  \hDbb
  =
  \hQbb\cdot
  \begin{pmatrix}
    \DX & 0 & 0 \\
    0 & \DY & \sqrt{\DY\Dr} \\
    0 & \sqrt{\DY\Dr} & \Dr
  \end{pmatrix}
  \cdot\hQbb^\T\,
  \label{eq:hDbb}
\end{equation}
with~$\DX = \nofrac{\EX}{\sigmaX^2}$, $\DY = \nofrac{\EY}{\sigmaY^2}$,
and~$\Dr = \nofrac{\EY\ell^2}{\sigmar^2}$.  The diffusion tensor couples
translational and rotational noises.  Our result is closely related to
\cite{Hottovy2012b}, but here the induced drift is due to angular dependence
rather than spatial inhomogeneity.

To go back and compare to the overdamped result
\cref{eq:overdamped_simplistic} obtained by simply neglecting the particle
mass, the Fokker--Planck equation~\eqref{eq:the_final_final_PDE} implies the
SDE
\begin{equation}
  \frac{\d}{\d\t}
  \begin{pmatrix}
    \xv \\ \phi
  \end{pmatrix}
  =
  \begin{pmatrix}
    \Uv+\Vv \\ \Omega
  \end{pmatrix}
  +
  \sqrt{2\hDbb}\cdot\dotwv
\end{equation}
where~$\sqrt{2\hDbb} = \hBbb^{-1}\cdot\hSigmabb$.  Note the additional
drift~$\Vv$.  The drift~$\Vv$ implies that the particle appears to swim at a
constant speed as in the ABP model~\eqref{eq:ABP_standard} for long times,
even for~$\Uv=0$.  The drift~$\Vv$ is only present when the fluctuating force
exerts a torque; it is an inertial effect that vanishes for isotropic
particles ($\sigmaX=\sigmaY$).  It does \emph{not} vanish for zero mass, since
it involves the ratio~$\Inerc/\mass$.

It is natural to form P\'eclet numbers based on the advective
time~$\aaa/\lvert\Vv\rvert$ and diffusive times~$\aaa^2/\DY$
and~$1/\Dr$, with~$\aaa$ the particle size:
\begin{align*}
  \PeY &= \frac{\lvert\Vv\rvert\aaa}{\DY} =
  \frac{2a\ell\sigmaY^2}{\sigmar}
  \frac{\lvert\sigmaX^{-1} - \sigmaY^{-1}\rvert}
  {1 + \Inerc\sigmaY/\mass\sigmar}\,
   \sim \frac{\ell}{\aaa},
  \\
  \Per &= \frac{\lvert\Vv\rvert}{\Dr\aaa}
  =
  \frac{2\sigmar}{\aaa\ell}\,
  \frac{\lvert\sigmaX^{-1} - \sigmaY^{-1}\rvert}
  {1 + \Inerc\sigmaY/\mass\sigmar}
  \sim
  \frac{\aaa}{\ell}.
\end{align*}
$\Pe_\perp$ is not large, but also not necessarily small.  $\Pe_\rot$ is a
dimensionless correlation length that diverges as~$\ell\rightarrow0$, since
the rotational diffusivity then vanishes.

We can compute the long-time effective diffusivity of the active particle.
Here there are two new effects: the noise-induced drift~$\Vv$ and the coupling
terms~$\sqrt{\DY\Dr}$ in the grand diffusion tensor~$\hDbb$.  Recall
that~$\hxv = (\xv,\phi)$, so $\hat{\xc}_3=\phi$.  The overdamped
Fokker--Planck equation~\eqref{eq:the_final_final_PDE} for~$\PP(\hxv,\t)$ is
\begin{multline}
  \pd_\t\PP
  + \Wc_i\,\pd_{\xc_i}\PP + \Omega\,\pd_\phi\PP
  =
  \pd_{\xc_i}\pd_{\xc_j}\l(\Dc_{ij}\,\PP\r) \\
  +
  2\pd_{\xc_i}\pd_\phi\bigl(\hDc_{i3}\,\PP\bigr)
  +
  \pd^2_\phi\bigl(\Dr\,\PP\bigr)
  \label{eq:FP_PP_explicit}
\end{multline}
where~$\Wv = \Uv + \Vv = \Wc\,\ruv$ is the total drift, and indices are summed
over~$1,2$.  To find the effective diffusivity, we
rescale~\eqref{eq:FP_PP_explicit} to focus on large
scales~$\delta^{-1} \sim \ell^{-1}$ and long times~$\delta^{-2}$,
with~$\delta$ a small parameter.  We
let~$\pd_\t \rightarrow \pd_\t + \delta^2\,\pd_\Time$,
and~$\pd_{\xv} \rightarrow \pd_{\xv} + \delta\,\pd_{\Xv}$ and expand
$\PP = \PPt(\Xv,\Time) + \delta\,\PP_1(\phi;\Xv,\Time) +
\delta^2\,\PP_2(\phi;\Xv,\Time) + \cdots$, where we anticipated the functional
dependencies to abridge the derivation.  (Our approach is equivalent to
\cite{Zia2010}, who average over angles, or \cite{Cates2013}, who expand~$\PP$
in harmonics.)  At order~$\delta^1$ we have
$\Dr\,\pd_\phi^2\PP_1 - \Omega\,\pd_\phi\PP_1 = \Wc_i\,\pd_{\Xc_i}\PPt -
2\pd_{\Xc_i}\pd_\phi\bigl(\hDc_{i3}\,\PPt\bigr)$, with a simple solution
linear in~$\cos\phi$ and~$\sin\phi$.  At order~$\delta^2$ we have
%\begin{multline}
%  \pd_\Time\PPt
%  + \Wc_i\,\pd_{\Xc_i}\PP_1
%  + \Omega\,\pd_\phi\PP_2
%  =
%  \pd_{\Xc_i}\pd_{\Xc_j}\l(\Dc_{ij}\,\PPt\r) \\
%  +
%  2\pd_{\Xc_i}\pd_\phi\bigl(\hDc_{i3}\,\PP_1\bigr)
%  +
%  \Dr\,\pd_\phi^2\PP_2
%  \label{eq:order2}
%\end{multline}
%with
the solvability condition
\begin{align}
  \pd_\Time\PPt
  &=
  \bigl\langle
  \Wc_i\,(\Wc_j - 2\pd_\phi\hDc_{j3})/\Dr \nonumber\\
  &\phantom{=\bigl\langle}
  +
  \Dc_{ij}
  \bigr\rangle\,
  \pd_{\Xc_i}\pd_{\Xc_j}\PPt
  \rdef
  \Deff\,\grad^2_{\Xv}\PPt
  \label{eq:longtime_heat}
\end{align}
where angle brackets are repurposed for angular averaging, and the effective
diffusivity is
\begin{subequations}
\begin{align}
  \Deff
  &=
  \tfrac12(\DX + \DY)
  +
  \Dt
  \label{eq:Deff} \\
  \Dt
  &\ldef
  \frac{\Wc\Dr}{2(\Dr^2 + \Omega^2)}
  \l(\Wc + \frac{2\EY\ell}{\sigmaY\sigmar}\r).
  \label{eq:Dt}
\end{align}
\end{subequations}
\Cref{eq:longtime_heat} displays the expected long-time isotropy of the
probability density.  Compare~$\Dt$ to~$\Uc^2/2\Dr$ for the ABP
model~\eqref{eq:ABP_standard} \cite{Howse2007, Peruani2007, Lindner2008,
  Golestanian2009, Fodor2016, Caprini2021}.

The new diffusivity~$\Dt$ combines contributions from the swimming~$\Uc$, the
noise-induced drift~$\Vv$, and from the coupling terms in~$\hDbb$.  From here
we set $\Uv=\Omega=\DX=0$ to highlight the new effects: the particle is
``shaking its hips'' but would be a non-swimmer if not for the noise-induced
drift; see also \cite{Thiffeault2022b} for a deterministic version.  (The
swimmer is a ``treadmiller'' or reciprocal swimmer that doesn't strictly swim,
but only diffuses~\cite{Crowdy2010,Lauga2011,Obuse2012}.)  In that case after
using~\eqref{eq:Vv} Eq.~\eqref{eq:Dt} becomes
\begin{align}
  \Dt_0
  =
  \frac{2\DY(1 + \Inerc\sigmaX/\mass\sigmar)}
  {(1 + \Inerc\sigmaY/\mass\sigmar)^2}
  \frac{\sigmaY}{\sigmaX}
  \l(\frac{\sigmaY}{\sigmaX} - 1\r).
  \label{eq:Ddrift}
\end{align}
The form~\eqref{eq:Ddrift} for~$\Dt_0$ has two striking features.  First, it
is negative for particles with $\sigmaY < \sigmaX$, so that it hinders
diffusion.  In fact, the combination~$\Dt_0 + \tfrac12\DY$ attains a minimum
of zero for~$\sigmaY = \sigmaX/(2 + \Inerc\sigmaX/\mass\sigmar)$.  A particle
satisfying this relation can only diffuse through~$\DX$ and thermal noise.

\begin{figure}
  \begin{center}
    \includegraphics[width=\columnwidth]{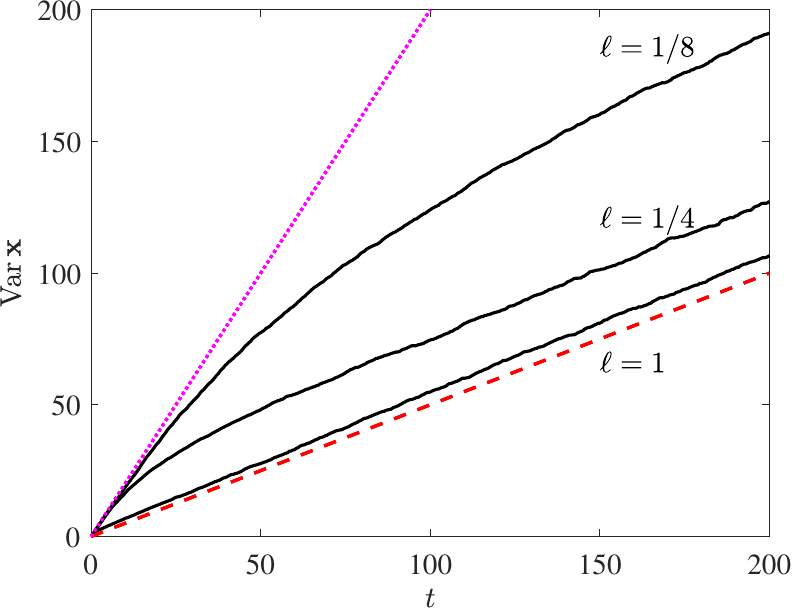}
  \end{center}
  \caption{The mean-squared displacement (variance) $\Var\xv$, averaged over
    $5\,000$ oblate non-swimming ($\Uv=\Omega=0$) active particles for various
    values of~$\ell$.  The upper dotted line is
    $4\times\tfrac12(\DX + \DY)\t$, and the bottom dashed line
    is~$4\Deff\t$.  As $\ell$ becomes smaller, there is a longer transient
    before the behavior begins to follow \cref{eq:longtime_heat}.  This
    transient diverges as $\ell \rightarrow 0$.  Parameter values:
    $\mass=\Inerc=.05$, $\sigmaX=2$, $\EY=\sigmaY=\sigmar=1$, $\EX=0$.}
  \label{fig:elldep}
\end{figure}

The second striking feature of~\eqref{eq:Ddrift} is that it is independent
of~$\ell$.  This is a paradox: for $\ell=0$, we have~$\Vv=0$
and~$\hDc_{i3}=0$, so none of the effects mentioned here occur.  The
resolution is that there is a transient of
duration~$\Dr^{-1}=\sigmar^2/\EY\ell^2 \sim \delta^{-2}$ before the long-time
form~\eqref{eq:longtime_heat} applies, and this transient becomes infinite
as~$\ell\rightarrow0$.  This transient can be seen in the simulations of the
full inertial equations~\eqref{eq:SDE_inertia} in \cref{fig:elldep},

It is important to note that the ratio~$\Dt_0/\DY$ is rarely negligible: all
the dimensionless ratios appearing on the right of \cref{eq:Ddrift} are
typically of order one.  The transient time scale~$\Dr^{-1}$ can be estimated
by~$\aaa^2/\DY$, where $\aaa$ is the particle size; if $\Dr^{-1}$ is very
long, then $\DY$ was likely negligible to begin with.  The modifications
discussed in this paper are thus likely to be relevant in many applications.

So why haven't these types of corrections been observed?  Many authors
simulate the ABP model directly, since the inertial
equations~\eqref{eq:SDE_inertia} are expensive to solve due the small step
size required, in which case the new effects are ruled out.  Particle
anisotropy is also seldom considered.  Experimentally, diffusivities are
measured directly from the distributions of displacements, and so any
connection between the rotational and translational diffusivities is typically
lost.  One approach might be to obtain the covariance matrix $\hAbb$ directly,
by measuring the correlations between translational and rotational velocities.
A nonzero correlation would indicate a coupling as predicted here.

In future work we will generalize the derivation to arbitrary
three-dimensional active particles\cite{Delong2015,Ilie2015}, with the
fluctuating force not necessarily applied on an axis of symmetry.  There are
several other possible extensions, such as the inclusion of multiple forces
and torques acting on the body.  The consequences to swim pressure
\cite{Takatori2014,Takatori2014b}, run-and-tumble dynamics
\cite{Subramanian2009,Cates2013}, non-Newtonian swimming \cite{Datt2019},
velocity-dependent friction~\cite{Erdmann2000}, and particle interactions
\cite{Fodor2016,Marath2019} also remain to be investigated.

\begin{acknowledgments}
  The authors thank
    Saverio Spagnolie,
    Hongfei Chen,
    Scott Hottovy,
    Christina Kurzthaler,
    Eric Lauga,
    Cesare Nardini,
    John Wettlaufer,
    and
    Ehud Yariv
  for helpful comments and discussions.
\end{acknowledgments}

\bibliography{journals_abbrev,needle}

\printnomenclature

\end{document}